\documentclass[11pt,a4paper,fleqn]{article} 
\catcode`\@=11
\@addtoreset{equation}{section}
\def\theequation{\arabic{section}.\arabic{equation}}
\def\appendix{\renewcommand{\thesection}{\Alph{section}}\setcounter{section}{0}
              \renewcommand{\theequation}
            {\mbox{\Alph{section}.\arabic{equation}}}\setcounter{equation}{0}}

\def\maketitle{\thispagestyle{empty}\setcounter{page}0\newpage
                \renewcommand{\thefootnote}{\arabic{footnote}}
                  \setcounter{footnote}0}
\renewcommand{\thanks}[1]{\renewcommand{\thefootnote}{\fnsymbol{footnote}}
               \footnote{#1}\renewcommand{\thefootnote}{\arabic{footnote}}}
\newcommand{\preprint}[1]{\hfill{\sl preprint - #1}\par\bigskip\par\rm}
\renewcommand{\title}[1]{\begin{center}\Large\bf #1\end{center}\rm\par\bigskip}
\renewcommand{\author}[1]{\begin{center}\Large #1\end{center}}
\newcommand{\address}[1]{\begin{center}\large #1\end{center}}
\newcommand{\pacs}[1]{\smallskip\noindent{\sl PACS numbers:
                       \hspace{0.3cm}#1}\par\bigskip\rm}
\def\babs{\hrule\par\begin{description}\item{Abstract: }\it} 
\def\eabs{\par\end{description}\hrule\par\medskip\rm}
\renewcommand{\date}[1]{\par\bigskip\par\sl\hfill #1\par\medskip\par\rm}

\def\dinfn{Dipartimento di Fisica, Universit\`a di Trento\\ 
                           and Istituto Nazionale di Fisica Nucleare,\\
                                   Gruppo Collegato di Trento, Italia \medskip}

\newcommand{\guido}{Guido Cognola\thanks{e-mail: \sl cognola@science.unitn.it\rm}}
\newcommand{\sergio}{Sergio Zerbini\thanks{e-mail: \sl zerbini@science.unitn.it\rm}}

\newcommand{\s}[1]{\section{#1}}

\def\hs{\qquad}               
\def\nn{\nonumber}            
\def\beq{\begin{eqnarray}}    
\def\eeq{\end{eqnarray}}      
\def\at{\left(}               
\def\aq{\left[}               
\def\ct{\right)}              
\def\cq{\right]}              
\def\R{{\hbox{{\rm I}\kern-.2em\hbox{\rm R}}}}   
\def\H{{\hbox{{\rm I}\kern-.2em\hbox{\rm H}}}}   
\def\N{{\hbox{{\rm I}\kern-.2em\hbox{\rm N}}}}   
\def\C{{\ \hbox{{\rm I}\kern-.6em\hbox{\bf C}}}} 
\def\Z{{\hbox{{\rm Z}\kern-.4em\hbox{\rm Z}}}}   
\def\ii{\infty}                                  
\newcommand{\fr}[2]{\mbox{$\frac{#1}{#2}$}}        
\def\Tr{\mathop{\rm Tr}\nolimits}                  
\renewcommand{\Re}{\mathop{\rm Re}\nolimits}       
\def\dir{/\kern-.7em D\,}                            
\def\lap{\Delta\,}                                 
\def\al{\alpha}
\def\be{\beta}
\def\ga{\gamma}
\def\de{\delta}
\def\ep{\varepsilon}
\def\ze{\zeta}

\def\la{\lambda}
\def\ro{\varrho}
\def\si{\sigma}

\def\Ga{\Gamma}

\def\La{\Lambda}

\def\Om{\Omega}

\begin{document}

\preprint{UTF 441}

\title{On the Dimensional Reduction Procedure}
\author{\guido and \sergio}
\address{\dinfn}

\date{February 2001}

\babs
The issue related to the so-called dimensional reduction procedure is 
revisited within the Euclidean formalism.
First, it is shown that for symmetric spaces, the local exact heat-kernel 
density is equal to
the reduced one, once the harmonic sum has been successfully performed.  
In the general case, due to the impossibility to deal with exact results, the  
 short $t$  heat-kernel asymptotics is considered.  It is found  that the 
exact heat-kernel and the dimensionally reduced one coincide up to two non 
trivial leading  contributions in
the short $t$  expansion. Implications of these results with regard to 
 dimensional-reduction anomaly are discussed.

\eabs



\s{Introduction}
\label{Form}

Very recently, motivated by applications concerning  the black hole physics 
initiated in \cite{wipf} and calculation of the effective action after and 
before the dimensional reduction \cite{eli},
Frolov, Sutton and Zelnikov  have introduced the concept of 
dimensional-reduction anomaly \cite{frolov99} related to a dimensional 
reduction procedure. Subsequently, the symmetric space cases
have been considered too \cite{sutton}. Very recently, 
consequences of the dimensional reduction anomaly  
in the Schwarzschild spacetime have also been analysed
\cite{balbinot00}.

To begin with, let us  recall  the general arguments leading to the 
proposal contained in the above mentioned papers.
Within the so-called one-loop approximation in Quantum 
Field Theory, the Euclidean one-loop effective action $\Ga$  may be 
expressed in terms of the
functional determinant of an elliptic differential operator $O$, defined on a 
$D$-dimensional manifold, namely
\beq
\Ga\sim\ln \det O\,. 
\eeq 
The ultraviolet
one-loop divergences, which  are present, may be regularised by means of 
 analytic regularisations, for example the zeta-function regularisation 
(for recent  reviews, see \cite{eliz94b,byts96-266-1}).

In the presence of space-times having  some symmetries like the 
D-dimensional space-time is the "warp" product 
$M_D=M_P \times \Sigma_Q$, $\Sigma_Q$ being a Q-dimensional 
symmetric space with constant curvature,  
the harmonic analysis on it is normally used in order to
 dimensionally 
"reduce" the relevant fluctuation operator $O$.
 It turns out that the 
quantum dimensional-reduced theory, defined by a specific procedure,  might 
be  not equivalent to the original 
one, and this fact is  related  to the presence of the 
dimensional-reduction anomaly.

The reason of this possible discrepancy has been  explained in \cite{frolov99}
as  mainly due to the necessity of the regularisation and renormalisation 
of the effective action in spaces with different dimensions. There, it has 
also been   observed a possible 
connection with  the so called multiplicative 
anomaly. Regarding this issue,    
see, for instance, 
\cite{eliz98-194-613,eliz98-57-7430,eliz98-532-407}.

Let us consider  a scalar field $\Phi$ propagating in the above 
mentioned space. Thus, one is mainly dealing with a second order 
self-adjoint non-negative differential operator
\beq
L_D=-\lap_D+m^2+\xi R_D\,, 
\label{lap1}\eeq
in which $\lap_D$ is the Laplace operator on $M_D$, $m^2$  
a possible mass term and $\xi R_D$ a suitable "potential term", 
describing the 
non-minimal coupling with the gravitational field. We will assume that the  
spectrum is
 bounded from below. The "exact" theory, namely the non-dimensional reduced 
one, may be 
described by the path integral (Euclidean partion function)
\beq
Z=\int D\Phi e^{-\int dV_D \Phi L_D \Phi}=e^{-\Ga}\,, 
\eeq
$dV_D$ being the infinitesimal volume on $M_D$.
The effective action $\Ga$ has to be regularised and may be  
expressed by means of a zeta-regularised functional determinant 
\cite{ray73-98-154,hawk77-55-133,dowk76-13-3224}
\beq
\Ga=-\ln Z=-\frac{1}{2}\aq \ze'(0|L_D)+\ln \mu^2 \ze(0|L_D)\cq\,,
\eeq
$\mu^2$ being the renormalisation parameter.
Here, the zeta-function is defined by 
\beq
\zeta(s|L_D)=\frac{1}{\Ga(s)}\int_0^\ii dt \ t^{s-1}K_t\,
\hs\hs K_t= \Tr e^{-t L_D}\,,
\label{mt}
\eeq
valid for $\Re s> D/2$. Here $ \Tr e^{-t L_D}=\sum_i  e^{-t \la_i}$, $\la_i$ 
being the eigenvalues of $L_D$. As a 
consequence, $ \ze(s|L_D)=\sum_i  \la_i^{-s}$.
If zero modes are present, one has to subtract them, replacing 
$\Tr e^{-t L_D}$ with $\Tr e^{-t L_D}-P_0$, $P_0$ being the projector onto the
zero modes. 

Of course one may use other regularisation procedures. 
As an example, the dimensional regularisation is defined by
\beq
\Ga_\ep&=&-\frac{1}{2}\int_0^\ii dt \ t^{\ep-1} \Tr e^{-t L_D/\mu^2}=
-\frac{1}{2}\Ga(\ep)\ze(\ep|L_D/\mu^2)\nn \\
&=&-\frac{1}{2}\aq \frac{\ze(0|L_D)}{\ep}+\ze'(0|L_D)
+(\ln\mu^2-\ga)\ze(0|L_D)+
O(\ep)\cq \,.
\label{mtdim}
\eeq
Other regularisations  may be used with $t^{\ep}$ substituted by a 
suitable regularisation function $\ro_\ep(t)$ 
(see, for example \cite{kirsten}).
Recall that the zeta-function regularisation is a {\it finite} regularisation
and corresponds to the choice
\beq
\ro_\ep(t)=\frac{d}{d\ep}\at\frac{t^{\ep}}{\Ga(\ep)}\ct\,.
\eeq
The other ones, as is clear from Eq. (\ref{mtdim}), 
give the same finite part, modulo a renormalisation, and contain divergent 
terms as the cutoff parameter $\ep \to 0$ and these divergent terms  have to 
be removed by related counter-terms.

As a consequence, as  will be shown in the following,  
a crucial role is played by the 
quantity $\Tr e^{-t L_D}$. With regard to this quantity, its  short-$t$ 
asymptotics  has been extensively studied.
For a second-order operator
on a boundary-less $D$-dimensional (smooth) manifold, it reads
\beq
K_t\simeq \sum_{j=0}^\ii A_j(L_D)\
t^{j-D/2}
\:,\label{tas0}
\eeq
in which $A_j(L_D)$ are the Seeley-DeWitt coefficients, which  can be computed
with different techniques \cite{dewi65b,seel67-10-172}. The divergent terms 
appearing in a generic 
regularisation depend on  $A_j(L_D)$.

In the sequel, we shall also deal with local quantities, which can be defined 
by the local zeta-function. With this regard, it is relevant the local short 
$t$ heat-kernel asymptotics, which reads (see Appendix B)
\beq
K_t(L_D)(x)=e^{-tL_D}(x)\simeq \frac{1}{(4 \pi )^{D/2}} 
\sum_{j=0}^\ii a_j(x|L_D)\
t^{j-D/2}
\:,\label{hkl}
\eeq
where $a_j(x|L_D) $ are the local Seeley-DeWitt coefficients. 
Some of them are well known and read 
\beq
a_0(x|L_D)&=& 1\,, \hs 
a_1(x|L_D)=\frac{R}{6}-X\,.\nn\\
a_2(x|L_D)&=&\frac{1}{2}[a_1(x|L_D)]^{2}+\frac{1}{6} \lap_D a_1(x|L_D)
\nn\\&&\hs\hs
+\frac{1}{180}\at\lap_D R+ R^{ijkr}R_{ijkr}-R^{ij}R_{ij}\ct\,,
\label{sde010}\eeq
where $X$ is a function which depends on the operator one is dealing with.
For example, for the Laplace-like operator in (\ref{lap1}),
$X=m^2+\xi\:R$. 

It may be convenient to re-sum partially this asymptotic expansion and one has
\cite{parker}
\beq
e^{-tL_D}(x)\simeq \frac{e^{t a_1(x|L_D)}}{(4 \pi)^{D/2}}
\sum_{j=0}^\ii b_j(x|L_D)\
t^{j-\fr{D}{2}}
\:.\label{hkl11}
\eeq
The advantage of the latter expansion with respect to the previous one, 
is due to the fact that now the $b_j$ coefficients depend 
on the "potential" $X$ only through its derivatives. 
In fact one has 
\beq
b_0(x|L_D)&=&1\,, \hs 
b_1(x|L_D)=0\,, \nn \\
b_2(x|L_D)&=&-\frac{1}{6}\lap_D a_1(x|L_D)+
\frac{1}{180}\at\lap_D R+ R^{ijkr}R_{ijkr}-R^{ij}R_{ij}\ct\,.
\label{sde01}
\eeq
Further coefficients  $b_j(x|L_D)$ are reported in Appendix B. 

The local zeta-function is defined by means of the Mellin transform, i.e.
\beq
\ze(s|L_D)(x)=\frac{1}{\Ga(s)}\int_0^\ii dt\:t^{s-1}e^{-tL_D}(x)\,.
\eeq 
Making use of the local zeta-function, one may evaluate the effective 
Lagrangian, which is proportional to $\ze'(0|L_D/\mu^2)(x)$, 
and the vacuum-expectation value of the quantum 
fluctuation given by (see for example\cite{moretti,binosi})
\beq
<\Phi(x)^2>=\lim_{\ep \to 0}\frac{d}{d\ep}
\aq \ep \mu^{2\ep}\ze(1+\ep|L_D)\cq\,,
\label{qf}
\eeq
which simplifies when $D$ is odd, due to the fact that in odd dimensions 
the zeta-function is regular at $s=1$ and so
\beq
<\Phi(x)^2>=\ze(1|L_D)\,.
\label{qfodd}
\eeq

Since the exact expression of the local zeta-function is known only in a 
limited number of cases, in general one has to make use of 
some approximations. If the coefficient $a_1(x|L_D)$ 
is very large and negative (this is 
true if the case of large mass), 
one may obtain an asymptotic expansion of the local 
zeta-function by means of the short $t$ expansion (\ref{hkl}) and the Mellin 
transform in the form \cite{kirsten}
\beq
\ze(s|L_D)(x)& \simeq& \frac{\Ga(s-\fr{D}{2})}{(4\pi )^{\fr{D}{2}}\Ga(s)}
\at -a_1(x|L_D)\ct^{\fr{D}{2}-s}\nn \\
&&\hs+\sum_{j=2}^\ii
\frac{\Ga(s+j-\fr{D}{2})}{(4\pi )^{\fr{D}{2}}\Ga(s)}
\at-a_1(x|L_D) \ct^{\fr{D}{2}-s-j}b_j(x|L_D)\,.
\label{voros}
\eeq 
The latter expansion directly gives the analytic 
continuation in the whole complex plane.
The global zeta-function can be obtained integrating over the manifold.

The content of the paper is the following. In Sec.~2, the 
dimensional-reduced theory is introduced and the 
formalism is developed. In Secs.~3 and 4, the two symmetric spaces  $R^D$ 
and H$^D$ are considered in some detail. In Sec.~5, the general
case is considered making use of the heat-kernel asymptotics. 
The paper ends with the conclusions and two Appendices, 
where  some technical details are reported. 

\section{Dimensional-reduced theory, dimensional reduced heat-kernel  and
 dimensional-reduction  anomaly}

Here we introduce the dimensional-reduced theory according to 
\cite{wipf,frolov99}.
We indicate by $\tilde{M}_D$ a $D$-dimensional Riemannian manifold  
with metric $\tilde g_{\mu\nu}$ and coordinates $\tilde x^\mu$ 
($\mu,\nu=1,...,D$)
and by ${M}_P$ and $\hat{M}_Q$ ($Q=D-P$) two sub-manifolds with 
coordinates $x^i$ ($i,j=1,...,P$) and $\hat x^a$ ($a,b=P+1,...,D$) and
metrics $g_{ij}$ and $\hat g_{ab}$ respectively, related 
to $\tilde g_{\mu\nu}$ by the warped product
\beq 
d\tilde s^2=\tilde g_{\mu\nu}d\tilde x^\mu d\tilde x^\nu=
g_{ij}(x)dx^idx^j+e^{-2\si(x)}\hat g_{ab}(\hat x)d\hat x^ad\hat x^b\:.
\label{metric}\eeq
Here,  $\hat{M}_Q=\Sigma_Q$ is a constant curvature symmetric space.  

We shall use the notation $\tilde R^\al_{\be\ga\de}$,
$R^i_{jmn}$ and $\hat R^a_{bcd}$ for Riemann tensors in 
$\tilde{M}_D$, ${M}_P$ and $\hat{M}_Q$ respectively, 
and similarly for all other quantities. 
In the Appendix A, one can find the relationship between the geometrical 
quantities related to the sub-manifolds.

We start with a scalar field $\Phi(\tilde x)$ in the Riemannian 
manifold $\tilde{M}_D$.
Using Eqs.~(\ref{lap}) and (\ref{RRR}) in the appendix~\ref{definizioni}, 
for the Laplacian-like operator we have
\beq 
L_D\Phi(\tilde x) =\tilde L\Phi(\tilde x)=(-\tilde\lap+\xi\tilde R+m^2)
\Phi(\tilde x)
=(L+e^{2\si}\hat L)\Phi(\tilde x)\:,
\eeq
where
\beq 
&&L=-\lap+Q\si^k\nabla_k+\xi\aq R+2Q\lap\si-Q(Q+1)\si^k\si_k\cq+m^2\:,
\\
&&\hat L=-\hat\lap \:.
\eeq
In order to dimensionally reduce the theory, let us introduce the harmonic
analysis on $\Sigma_Q$ by means of 
\beq
\hat LY_\al(\hat x)=\la_\al Y_\al(\hat x)\:,
\eeq
$\la_\al,Y_\al$ being the eigenvalues and eigenfunctions 
of $\hat L$ on the symmetric space  $\Sigma_Q=\hat{M}_Q$.
For any scalar field in $\tilde{M}_D$, we can write
\beq 
\Phi(\tilde x)=\sum_\al\phi_\al(x)Y_\al(\hat x)
\eeq
and for the partition function, after integration over $Y_\al$ in 
the classical 
action,
\beq 
 Z^*&=&\int\:d[\bar\phi]e^{-\int\hat\phi\tilde L\hat\phi\:
\sqrt{\tilde g}\:d^Pxd^Q\hat x}=\prod_\al Z_\al\:,
\eeq
where 
\beq 
Z_\al=\int\:d[\bar\phi_\al]e^{-\int\bar\phi_\al L_\al\bar\phi_\al\:d^Px\:}\:.
\eeq
Here $\bar\phi=\sqrt[4]{\tilde g}\phi$ and $\bar\phi_\al=\sqrt[4]{g}\phi_\al$ 
are scalar densities of weight $-1/2$ and the dimensional reduced operators
read
\beq 
&&L_\al=-\lap+V+e^{2\si}\la_\al\:,
\label{Lal}\nn\\
&&V=m^2+\xi\aq R+2Q\lap\si-Q(Q+1)\si^k\si_k\cq
-\frac{Q}2\lap\si+\frac{Q^2}{4}\si^k\si_k\:.
\label{V}\eeq 
In the following, we will denote by an asterix all the quantities associated 
with 
the dimensional reduced operators. As a result, we formally have
\beq
Z^*=\prod_\al \at \det\frac{L_\al}{\mu^2} \ct^{-1/2}\,.
\eeq
If we {\it ignore} the multiplicative anomaly associated with functional 
determinants, we have  

\beq
\Ga^*=-\ln Z^*=\frac{1}{2}\sum_\alpha \ln \det\frac{ L_\alpha}{\mu^2}\,.
\label{redac=}
\eeq
This formal expression  may be regularised and renormalized and we have
\beq
\Ga^*_\ep=-\frac{\mu^{2\ep}}{2}\sum_\al \int_0^\ii dt t^{-1} g_\ep(t)
\Tr e^{-tL_\al}\,.
\eeq
Removing the cutoff and, for example  making use of a finite regularisation,
one arrives at
\beq
\Ga^*=\frac{1}{2}\sum_\alpha \zeta'(0|\frac{ L_\alpha}{\mu^2})\,.
\label{redac==}
\eeq
Within this procedure, a quite natural  definition of the  
dimensional-reduction anomaly is \cite{frolov99} 
\beq
A_{DRA}= \Ga-\Ga^* \,.
\label{redeff}
\eeq
However, there exists {\it another} possible procedure: 
 if we {\it do not remove} the ultraviolet cutoff $\ep$, 
we may interchange 
the harmonic sum and the integral and arrive at
\beq
\Ga^*_\ep=-\frac{\mu^{2\ep}}{2} \int_0^\ii dt\:t^{-1}g_\ep(t)
 K_t^*\,,
\label{redhk*}
\eeq
where we have introduced the dimensionally reduced heat-kernel trace  
\beq
K_t^*=\sum_\alpha \Tr e^{-t L_\alpha}\,. 
\label{redheat}
\eeq
It is clear that within  this second procedure, the existence of a non 
vanishing 
dimensional 
reduction anomaly is strictly related to the fact whether the identity
\beq 
K_t^*=K_t
\label{=kt}
\eeq
holds. 

In the following Sections, the validity of the identity
(\ref{=kt}) will be investigated.

\section{The flat symmetric space $R^D$}

In this Section we shall show that, by performing exactly the harmonic sum,
the  Eq. (\ref{=kt}) holds. 

Let us consider a free massive scalar field in 
the D-dimensional Euclidean space $R^D$. 
The relevant operator is
\beq
L_D=-\lap_D+m^2
\eeq
and the corresponding heat-kernel reads
\beq
K_t(L_D)=\frac{e^{-t m^2}}{(4 \pi t)^{D/2}}\:V(R^D)\:,
\hs\hs V(R^D)=\int d\tau\int d\vec x\:.
\label{mhk}\eeq
Here $V(R^D)$ is the (infinite) volume of the manifold. 

We can also use spherical co-ordinates. Then
\beq
ds^2=d\tau^2+dr^2+r^2 dS^2_Q\,,
\eeq
where $dS^2_Q$ is the measure of the $Q=(D-2)$-dimensional unitary sphere
$S_Q$, which plays the role of the symmetric space $\hat M_Q$ in
the previous  section.
In spherical co-ordinates, the volume is given by
\beq 
V(R^D)=\Om_Q\:\int\:d\tau\int\:r^Q\:dr\:,
\hs\hs \Om_Q=\frac{2\pi^{\fr{Q+1}{2}}}{\Ga(\fr{Q+1}{2})}\:,
\nn\eeq 
$\Om_Q$ being the volume of the sphere $S_Q$.
The warp factor $e^{-2\si}=r^2$ and $V$, in Eq.~(\ref{V}), becomes
\beq 
V=m^2+\frac{Q}{2}\at\frac{Q}{2}-1\ct\frac{1}{r^2}\:.
\nn\eeq

The spectrum of the Laplacian $\hat\lap$ on the sphere $S_Q$
is well known. In fact, the eigenvalues $\la_\al\equiv\la_l$ 
and the corresponding degeneration are given by
\beq
\la_l=l(l+Q-1)\,,\hs\hs l\geq0\,,
\eeq
\beq
D_l^Q=(2l+Q-1)\:\frac{(l+Q-2)!}{(Q-1)!\:l!}\,,
\hs\hs D_0^Q=1\:.
\eeq
From Eq.~(\ref{Lal}) then we directly have
\beq
L_l=-\partial^2_\tau-\partial^2_r+\frac{C_l}{r^2}+m^2\,,
\hs C_l=\frac{(2l+Q-1)^2-1}4.
\eeq
Using the factorisation property of the heat kernel \cite{byts96-266-1},
for the operator $L_l$ we get
\beq 
K_t(L_l)=K_t\at-\partial^2_r+\frac{C_l}{r^2}\ct K_t(-\partial^2_\tau+m^2)
=K_t\at-\partial^2_r+\frac{C_l}{r^2}\ct\:
\frac{e^{-tm^2}}{\sqrt{4\pi t}}\:\int\:d\tau\:.
\nn\eeq
The spectrum of $-\partial^2_r+C_l/r^2$ is continuous and non negative.
We indicate by $\la^2$ the eigenvalues
and by $\psi_l$ the eigenfunctions which are given by
\beq 
\psi_l(r)=\sqrt{r}\:J_{\nu_l}(\la r)\:,
\hs\hs \nu_l=l+\frac{Q-1}2\:,
\nn\eeq
$J_\nu$ being the Bessel functions of the the first kind. 
The heat-kernel density, by definition, is given by
\beq 
K_t(r|L_l)=\int_{0}^{\ii}\:e^{-t\la^2}|\psi_l(r)|^2\:\la\:d\la
=\frac{re^{-r^2/2t}}{2t}\:I_{\nu_l}\at\frac{r^2}{2t}\ct\:.
\nn\eeq
For the whole operator we finally get 
\beq
K^*_t(L_D)=\frac{e^{-tm^2}}{\sqrt{4\pi t}}\:\int\:d\tau\:
\int\:\frac{re^{-r^2/2t}}{2t}\sum_{l=0}^\ii D_l^Q 
I_{\nu_l}\at\frac{r^2}{2t}\ct\:dr\,.
\label{Kts}\eeq
This must be compared with Eq.~(\ref{mhk}). 
To this aim we recall some properties of the modified Bessel functions
$I_\nu$  and the Legendre polynomials, that is
\beq
\sum_{n=-\ii}^\ii\: t^{n}I_n(z)=e^{\frac{z(t^2+1)}{2t}}\:,\hs\hs
I_n(z)=I_{-n}(z)\:,\hs\hs n=1,2,...
\eeq
\beq 
2\nu I_\nu(z)=z\aq I_{\nu-1}(z)-I_{\nu+1}(z)\cq\:,
\nn\eeq
\beq
\sum^\ii_{l=0}(2l+1)P_l(t)I_{l+1/2}(z)=
\sqrt{\frac{2z}{\pi}}\:e^{zt}\,,\hs
P_l(1)=1\:.
\eeq
From the latter equations with $t=1$ we obtain
\beq 
\sum_{l=0}^\ii D_l^1\:I_{l}(z)=\sum_{n=-\ii}^\ii\:I_n(z)=e^z\:,
\label{bess1}\eeq
\beq 
\sum_{l=0}^\ii D_l^2\:I_{l+1/2}(z)=
\sum^\ii_{l=0}(2l+1)I_{l+1/2}(z)=
\sqrt{\frac{2z}{\pi}}\:e^{z}\,.
\label{bess2}\eeq

Now, by induction, we are able to prove the following equation:
\beq 
\sum_{l=0}^\ii D_l^Q I_{\nu_l}(z)=\frac{1}{\Ga(2\nu_0+1)}\:
\sum_{\nu=\nu_0}^\ii\frac{\Ga(\nu+\nu_0)}{\Ga(\nu+1-\nu_0)}\:
2\nu I_{\nu}(z)=
\frac{z^{\nu_0}\:e^z}{2^{\nu_0}\Ga(\nu_0+1)}\:,
\label{iden}\eeq
where the sum is over integer or half-integer $\nu$ according
to whether $\nu_0=(Q-1)/2$ is integer or half-integer.

For $Q=1$ ($\nu_0=0$) and $Q=2$ ($\nu_0=1/2$), the latter equation 
reduces to (\ref{bess1}) and (\ref{bess2}) respectively. Then,
let us suppose Eq.~(\ref{iden}) to be true for a given $\nu_0$ and
show that it remains true also for $\nu_0+1$. In fact we have
\beq 
&&\frac{1}{\Ga(2\nu_0+3)}
\sum_{\nu=\nu_0+1}^\ii\frac{\Ga(\nu+\nu_0+1)}{\Ga(\nu-\nu_0)}\:
2\nu I_{\nu}(z)
=\nn\\&&
\frac{z}{\Ga(2\nu_0+3)}
\sum_{\nu=\nu_0+1}^\ii\frac{\Ga(\nu+\nu_0+1)}{\Ga(\nu-\nu_0)}\:
\aq I_{\nu-1}(z)-I_{\nu+1}(z)\cq
=\nn\\&&
\frac{z}{\Ga(2\nu_0+3)}
\sum_{\nu=\nu_0}^\ii\frac{\Ga(\nu+\nu_0)}{\Ga(\nu-\nu_0+1)}\:I_\nu(z)
\aq(\nu+\nu_0)(\nu+\nu_0+1)-(\nu-\nu_0)(\nu-\nu_0-1)\cq
=\nn\\&&
\frac{z(2\mu_0+1)}{2(\nu_0+1)}\frac{1}{\Ga(2\nu_0+1)}
\sum_{\nu=\nu_0}^\ii\frac{\Ga(\nu+\nu_0}{\Ga(\nu-\nu_0+1)}\:2\nu\:I_\nu(z)
=\nn\\&&
\frac{z^{\nu_0+1}\:e^z}{2^{\nu_0+1}\Ga(\nu_0+2)}\:.
\nn\eeq

Using (\ref{iden}) in (\ref{Kts}) we finally obtain the 
announced result 
\beq
K^*_t(L_D)=\frac{e^{-tm^2}}{(4\pi t)^{D/2}}\:\Om_Q\:
\int\:d\tau\int\:r^Q\:dr=K_t(L_D)\:.
\eeq

We conclude this Section observing that the above result also 
holds for an Euclidean Rindler space, since it  
can be regarded as the product of an Euclidean space in polar
coordinates times an Euclidean transverse space.

\section{The non-flat symmetric spaces $H^D$} 

In this Section, we complete the analysis on the validity  of the 
Eq. (\ref{=kt})  considering the case $H^D$. For the sake of simplicity,
we shall not deal with $S^D$, which is  more involved and can be found in 
\cite{camporesi}. 

It is convenient to make use of the Poincar\'e form for the metric of $H^D$,
namely
\beq
ds^2=\frac{1}{x^2}\at d^2\vec{y}+ dx^2 \ct\,.
\eeq
Thus, we have the warp product $R\times R^Q$ ($P=1$, $Q=D-1$). 
The manifold is non compact and the spectrum of $L_D=-\lap_D$  is 
continuous.
The diagonal part of the heat-kernel of $L_D$ on $H^D$ can be obtained 
directly making 
use of the Selberg transform (see, for example, \cite{byts96-266-1} 
and references  cited there) and reads    
\beq
K_t(L_D)=V_D\int_0^\ii d\la e^{-t \la^2}\Phi_D(\la)\,,
\label{hc}
\eeq
where $V_D$ is the volume of $H^D$ and
$\Phi_D(\la)$ the Harish-Chandra-Plancherel measure given by
\beq
\Phi_D(\la)=\frac{2}{(4\pi)^{D/2}\Ga(D/2)}
\frac{|\Ga(i\la+(D-1)/2)|^2}{|\Ga(i\la)|^2}
\:.\label{Plan}
\eeq

Now let us start the computation of $K^*_t(L_D)$. In the Poincar\'e 
representation, $\hat M_Q=R^Q$ is again a non compact manifold and the 
spectrum of the Laplacian $-\hat\lap_Q$ on it is continuous and formed by 
$\la_\alpha\equiv\la_{\vec k}=\vec k^2 >0$. 
The reduced harmonic operator is
\beq
L_\alpha=L_{\vec k}=-x^2\partial^2_x+(Q-1)x \partial_x+ \vec k^2 x^2\,.  
\eeq
The related heat-kernel can be computed by means of the Harish-Chandra method
\cite{bcz96}, namely by solving the generalized eigenfunction equation
\beq
\aq -x^2 \partial^2_x +(Q-1)x \partial_x + \vec k^2 x^2 \cq f_\la(x)=
\la^2 f_\la(x)\,.  
\eeq
The only solutions with the corrected behavior at the infinity are
\beq
f_\la(x)=x^{Q/2}K_{i\la}(xk)\,, \hs k=|\vec k|\,, 
\eeq
where $K_z(x)$ is the modified Bessel function.
The spectral theorem gives for the local reduced heat-kernel
\beq
K_t(x|L_{\vec k})=\frac{x^Q}{(2\pi)^Q}\int_0^\ii d\la \mu(\la) e^{-t\la^2}
K_{i\la}^2(xk)\,,
\eeq
where $\mu(\la)$ is the Kontarevich measure, namely
\beq
\mu(\la)=\frac{2}{\pi|\Ga(i\la)|^2}=\frac2{\pi^2}\,\la\sinh\pi \la\,.
\eeq
The harmonic sum is now   replaced by an  integral and we have
\beq
K^*_t(x|L_D)=\int_{R^Q} d\vec k K_t(L_{\vec k})=\frac{x^{Q}\Omega_{Q-1}}
{(2\pi)^Q}\int_0^\ii d\la \mu(\la) 
e^{-t\la^2}
\int_0^\ii dk k^{Q-1}K_{i\la}^2(xk)\,. 
\eeq
The integration over $k$ involving the modified Bessel function can be done
as well as the trivial integration over the coordinates and the result is
\beq
K^*_t(L_D)=V_D\frac{2}{(4\pi)^{D/2}\Ga(D/2)}\int_0^\ii d\la 
\frac{|\Ga(i\la+(D-1)/2)|^2}{|\Ga(i\la)|^2}  e^{-t\la^2}\,.
\eeq
Recalling Eqs.~(\ref{hc}) and (\ref{Plan}), one finally gets
\beq
K^*_t(L_D)=K_t(L_D)\,.
\eeq
As a consequence, also in the  constant curvature space
$H^D$, Eq. (\ref{=kt}) holds. 

\section{The general case} 

The result of the previous Sections tell us that it is very crucial to be 
able to
perform the harmonic sum. In general, this is not possible. For this 
reason,  we shall restrict ourselves to the class of 
non-trivial 
warped space-time of the kind considered in Sec.~2 and make use of the 
short $t$ heat-kernel expansion. For the exact theory we have (here 
$L_D=\tilde L$)   
\beq 
K_t(\tilde L)=\Tr e^{-t \tilde{L}}&\sim& \frac1{(4\pi t)^{D/2}}
\sum_{n=0}^\ii\:\tilde a_n(\tilde x|\tilde L)t^n\:,
\eeq
with
\beq 
\tilde a_1&=&a_1+e^{2\si}\hat a_1
-\frac{Q}6\aq\lap\si-\at\frac{Q}2-1\ct\si^k\si_k\cq\:,
\label{ta1}\\
\tilde a_2&=&a_2+e^{4\si}\hat a_2+e^{2\si}a_1\hat a_1
-\frac1{90}\si^k\nabla_k R-\frac1{45}...
\label{ta2}\eeq
where, as in Sec.~2, all quantities with tilde refers to the whole
manifold $\tilde{M}_D$ and all quantities with hat refers to the
sub-manifold $\hat{M}_Q$.  

With regard to the dimensional reduced kernel 
\beq 
K_t^*(\tilde L)&=&\sum_\al\Tr e^{-tL_\al}\:,\label{KtLal}
\eeq
where
\beq 
&&L_\al=-\lap+V+e^{2\si}\la_\al\:,
\nn\\
&&V=m^2+\xi\aq R+2Q\lap\si-Q(Q+1)\si^k\si_k\cq
-\frac{Q}2\lap\si+\frac{Q^2}{4}\si^k\si_k\:,
\eeq 
the short $t$ expansion can be computed by means of a  
straightforward computation, which is summarized in the following. 
Recalling the heat-kernel expansion relations 
(see Appendix~\ref{heat}), we can write
\beq 
K_t^*(\tilde L)&\sim&\sum_\al\frac{1}{(4\pi t)^{D/2}}
\int_{\tilde{M}_D}\:\sqrt{\tilde g}d^D\tilde x\:
e^{t(a_1-e^{2\si}\la_\al)}
\sum_{n=0}^\ii\:b_n(x|L_\al)t^n\nn\\
&=&\sum_\al\frac{1}{(4\pi t)^{D/2}}
\int_{\tilde{M}_D}\:\sqrt{\tilde g}d^D\tilde x\:e^{t(a_1-e^{2\si}\la_\al)}
\nn\\&&\hs\hs\hs\times\:
\aq1+\sum_{n\geq2;0\leq k\leq 2n/3}\:\La_{nk}\la_\al^kt^n\cq\:.
\label{Kt}\eeq
Now we observe that 
\beq 
K^*_\tau(\hat L)&=&\frac{1}{(4\pi\tau)^{Q/2}}
\int_{\hat{M}_Q}\:\sqrt{\hat g}d^Q\hat x\:
\sum_\al e^{-\tau\la_\al}\sim\frac{1}{(4\pi\tau)^{Q/2}}
\int_{\hat{M}_Q}\:\sqrt{\hat g}d^Q\hat x\:
\sum_{l=0}^\ii \hat a_l\tau^l\:,
\eeq
\beq
\frac{d^k K_\tau^*(\hat L)}{d\tau^k}&=&\frac{(-1)^k}{(4\pi\tau)^{Q/2}}
\int_{\hat{M}_Q}\:\sqrt{\hat g}d^Q\hat x\:
\sum_\al e^{-\tau\la_\al}\la_\al\nn\\
&\sim&\frac{1}{(4\pi\tau)^{Q/2}}
\int_{\hat{M}_Q}\:\sqrt{\hat g}d^Q\hat x\:
\sum_{l=0}^\ii\:\frac{\Ga(l-Q/2+1)}{\Ga(l-Q/2+1-k)}\:
\hat a_l\tau^{l-k}\:.
\label{Hder}\eeq
By setting $\tau=e^{2\si}$ and using Eq.~(\ref{Hder}) in Eq.~(\ref{Kt}), 
we finally obtain
\beq 
K_t^*(\tilde L)&\sim&\frac{1}{(4\pi t)^{D/2}}
\int_{\tilde{M}_D}\:\sqrt{\tilde g}d^D\tilde x\:
\sum_{l,j=0}^\ii
\frac{a_1^j\hat a_lt^{l+j}}{j!}\:e^{2l\si}
\nn\\&&\times\:
\aq1+\sum_{n\geq2;0\leq k\leq 2n/3}(-1)^k
\frac{\Ga(l-Q/2+1)}{\Ga(l-Q/2+1-k)}
\La_{nk}e^{-2k\si}t^{n-k}\cq 
\nn\\&&\sim  \frac1{(4\pi t)^{D/2}}
\sum_{n=0}^\ii\:\tilde a^*_n(\tilde x|\tilde L)t^n
\:.
\label{KtExp}\eeq

By the latter equation we immediately read off the heat kernel coefficients.
In particular, we get
\beq 
 a^*_1(\tilde x|\tilde L) &=&a_1+e^{2\si}\hat a_1+\frac{Q}2e^{-2\si}\La_{21}+ 
\frac{Q}4(Q+2)e^{-4\si}\La_{32}\:,
\label{c1}\eeq
\beq
a^*_2(\tilde x|\tilde L)&=&\frac12a_1^2+e^{4\si}\hat a_2
+e^{2\si}a_1\hat a_1+\La_{20}
+\frac{Q}2a_1e^{-2\si}\La_{21}
\nn\\&&\hs
+\frac{Q}4(Q-2)\hat a_1\La_{21}
+\frac{Q}2e^{-2\si}\La_{31}
+\frac{Q}4(Q+2)a_1e^{-4\si}\La_{32}
\nn\\&&
+\frac{Q}4(Q-2)\hat a_1e^{-2\si}\La_{32}
+\frac{Q}4(Q+2)e^{-4\si}\La_{42}
+\frac{Q}8(Q+2)(Q+4)e^{-6\si}\La_{53}
\nn\\&&\hs
+\frac{Q}{16}(Q+2)(Q+4)(Q+6)e^{-8\si}\La_{64}\:,
\label{c2}\eeq

Using the relations in Appendices~\ref{definizioni} and \ref{heat} 
one can show that the latter coefficients $ a^*_1(\tilde x|\tilde L)$ and
 $ a^*_2(\tilde x|\tilde L)$ 
exactly coincide with $\tilde a_1$ and $\tilde a_2$,
Eqs.~(\ref{ta1}), (\ref{ta2}).

As a consequence, one has
\beq
K_t(\tilde{L})\simeq 
\frac{e^{t \tilde{a}_1(\tilde x|\tilde L)}}{(4 \pi t)^{D/2}}
\aq 1+ b_2(\tilde x|\tilde L)t^2+ b_3(\tilde x|\tilde L)t^3+... \cq
\:,\label{hklp}
\eeq
\beq
K_t^*(\tilde{L})\simeq 
\frac{e^{t \tilde{a}_1(\tilde x|\tilde L)}}{(4 \pi t)^{D/2}}
\aq 1+ b_2(\tilde x|\tilde L)t^2+ b_3^*(\tilde x|\tilde L)t^3+... \cq
\:.\label{hklp*}
\eeq

What about $b_3^*(\tilde x|\tilde L)$,  $b_4^*(\tilde x|\tilde L)$, ...? 
Within our short $t$ approximation, we 
are not able to say anything about the relationship   with 
 $b_3(\tilde x|\tilde L)$,
 $b_4(\tilde x|\tilde L)$, ...
However, it is quite natural to  make the  conjecture that  
$b_n^*(\tilde x|\tilde L)=b_n(\tilde x|\tilde L)$ for every $n$ and Eq. 
(\ref{=kt}) holds exactly.

\section{Conclusions.}

In this paper, the issue related to the dimensional reduction 
procedure  has been revisited.
In the symmetric and constant curvature 
space-times, as $R^D$ and $H^D$, we have shown that the two local 
diagonal heat-kernels, namely the exact one and the one obtained summing  
 the dimensional reduced harmonic heat-kernels, are equal. This result 
holds for a generic symmetric space. 

In the general case, due to the impossibility to deal with exact quantities,
we have used a short $t$  expansion and a partial re-summation 
of the heat-kernel expansion. We have conjectured that the exact and 
total dimensional reduced  kernel are equal. Let us discuss about the 
consequences of this statement. 

After the dimensional reduction, as far as the
effective action is concerned, the 
operation of renormalisation
(addition of counter-terms and remotion of the cutoff) 
and the evaluation of the harmonic sum 
{\it do not} commute. If we keep fixed and non vanishing   the 
regularisation parameter, we may perform the harmonic sum, and if 
(\ref{=kt}) holds,  
we  may reconstruct the exact partition function, after renormalisation. 
In such a case, it is evident that no dimensional reduction anomaly occurs. 

On the other hand, one may remove the cutoff, adding the necessary 
counter-terms  or using a finite regularisation like the zeta-function
and perform the harmonic sum at the end. 
In this case, as shown in reference \cite{frolov99}, one 
has to correct the result by adding dimensional 
reduction anomaly terms.

It has to be stressed that the appearance of the reduction anomaly
is independent on the regularisation scheme one is dealing with,
since the regularised effective action, after the removal of 
divergences, is the same for all regularisations
\cite{kirsten}.

There exists also a mathematical reason for the necessity of these reduction 
anomaly terms. In fact, the harmonic sum of the renormalized dimensionally 
reduced effective action diverges and the dimensional anomaly reduction terms
are also  necessary to recover the exact and {\it finite} result.     
This is a consequence of the following asymptotic behaviour of the 
partial renormalized effective action, valid for very large $\la_\al$
(for the sake of simplicity here $D$ is assumed to be odd)
\beq
\ze'(0|L_\al)\simeq \Ga(-D/2)(V+\la_\al e^{2\si})^{D/2}+...\,.
\eeq
As a result, the harmonic sum over $\al$ is badly divergent!
This fact stems  also from the necessity of  the presence of the 
multiplicative 
anomaly, since it also diverges, 
being associated with a product of an  
infinite number of dimensional reduced operators 
\cite{eliz98-194-613,eliz98-532-407}.  

As a consequence, any approximation \cite{all} based on the truncation in the 
harmonic 
sum of the dimensional reduced theory, may lead, with regard to the 
comparison with the exact theory,  to incorrect conclusions 
(see also the discussions 
and further references reported in \cite{all1}).

\appendix


\s{Relations between curvatures}
\label{definizioni}
Here we write down the relations between curvatures and Laplacian
on the manifolds we are dealing with.

For the non vanishing components of connections and curvatures we directly 
obtain
\beq 
\tilde\Gamma^k_{ij}&=&\Gamma^k_{ij}\:,\hs\hs
\tilde\Gamma^k_{ab}=e^{-2\si}\si^k \hat g_{ab}\:,
\nn\\
\tilde\Gamma^a_{bc}&=&\hat\Gamma^a_{bc}\:,\hs\hs
\tilde\Gamma^a_{kb}=-\si_k\de^a_b\:,
\nn\eeq
\beq 
\tilde R_{ijmn}&=&R_{ijmn}\:,\nn\\
\tilde R_{abcd}&=&e^{-2\si}\hat R_{abcd}
-e^{-4\si}\at\hat g_{ac}\hat g_{bd}-\hat g_{ad}\hat g_{bc}\ct\:,\nn\\
\tilde R_{iajb}&=&e^{-2\si}\at\si_{ij}-\si_i\si_j\ct\:,
\nn\eeq
where
\beq 
\si_k=\nabla_k\si(x)\:,\hs\si^k=g^{kj}\si_j\:,
\hs\si_{ij}=\nabla_i\nabla_j\si(x)\:,
\nn\eeq
are the covariant derivatives, in the metric $g$, of the scalar function
$\si(x)$.
By contraction we get
\beq 
\tilde R_{ij}&=&R_{ij}+Q\at\si_{ij}-\si_i\si_j\ct\:,\nn\\
\tilde R_{ab}&=&\hat R_{ab}+e^{-2\si}\hat g_{ab}
\at\lap\si-Q\si^k\si_k\ct\:,
\nn\eeq
\beq
\tilde R=R+e^{2\si}\hat R+2Q\lap\si-Q(Q+1)\si^k\si_k\:,
\label{RRR}\eeq
and also
\beq 
\tilde R^{\al\be\ga\de}\tilde R_{\al\be\ga\de}&=&
R^{ijmn}R_{ijmn}+e^{4\si}\hat R^{abcd}\hat R_{abcd}
-4e^{2\si}\hat R\at\si^k\si_k\ct
\nn\\&&\hs
+4Q\si^{ij}\si_{ij}-8Q\si^{ij}\si_i\si_j
+2+Q(Q+1)(\si^k\si_k)^2\:,
\eeq
\beq 
\tilde R^{\al\be}\tilde R_{\al\be}&=&
R^{ij}R_{ij}+e^{4\si}\hat R^{ab}\hat R_{ab}+
2QR^{ij}\at\si_{ij}-\si_i\si_j\ct\nn\\
&&\hs+2e^{2\si}\hat R\at\lap si-Q\si^k\si_k\ct
+Q^2\si^{ij}\si_{ij}-2Q^2\si^{ij}\si_i\si_j
\nn\\&&\hs\hs
+Q(\lap\si)^2-2Q^2\si^k\si_k\lap\si+Q^2(Q+1)(\si^k\si_k)^2\:.
\eeq
Finally, for the Laplacian of any scalar function 
$f(\tilde x)$ on $\tilde{M}_D$ we have
\beq 
\tilde\lap f(\tilde x)
=\at\lap+e^{2\si}\hat\lap-Q\si^k\nabla_k\ct f(\tilde x)\:.
\label{lap}\eeq


\s{Heat kernel coefficients}
\label{heat}

The heat kernel for a Laplacian-like operator $L=-\lap+X$ on 
a $P$-dimensional curved manifold without boundary 
is usually written in the form
\beq 
K_t(x|L)\sim \at4\pi t\ct^{-P/2}
\sum_{n=1}^\ii\: a_n t^{n/2}\:,
\eeq
where the spectral coefficients $a_n$ are computable quantities depending
on $V$, its covariant derivatives and all geometric invariants.   
There exists also an alternative expansion \cite{parker} 
which has the advantage with respect to the previous one that the
expansion coefficients depend on $V$ only by its covariant derivatives.
It reads
\beq 
K_t(x|L)\sim \frac{e^{ta_1}}{\at4\pi t\ct^{P/2}}
\sum_{n=0}^\ii\: b_n t^{n/2}\:.
\eeq
Some coefficients are explicitly known and can be found 
in Refs.~\cite{parker}. They read
($a_0=b_0=1$ by definition),
\beq 
a_1=\frac{R}{6}-X\:,\hs\hs b_1=0\:,
\eeq
\beq
b_2&=&\frac16\:\lap a_1+\frac{1}{180}
\at\lap R+R^{ijrs}R_{ijrs}-R^{ij}R_{ij}\ct\:,
\\
b_3&=&\frac1{12}\nabla^kX\nabla_kX-\frac1{60}\lap^2X
-\frac1{90}R^{ij}\nabla_i\nabla_jX
-\frac1{30}\nabla^kR\nabla_kX+...\:,
\\
b_4&=&\frac1{72}(\lap X)^2
+\frac1{90}\nabla^i\nabla^jX\nabla_i\nabla_jX
\nn\\&&\hs
+\frac1{30}\nabla^kX\nabla_k\lap X
+\frac{1}{60}R^{ij}\nabla_iX\nabla_jX+...\:,
\\
b_5&=&-\frac1{72}\nabla^kX\nabla_kX\lap X
-\frac1{60}\nabla^iX\nabla^jX\nabla_i\nabla_j X+...\:,
\\
b_6&=&\frac1{288}\at\nabla^kX\nabla_kX\ct^2+...\:,
\eeq
where $...$ stand for lower terms in $X$.

The operator we are dealing with in the paper is
\beq 
L_\al=-\lap+V+e^{2\si}\la_\al
\nn\eeq
and so we write
\beq 
b_n=\sum_{0\leq k\leq2n/3}\La_{nk}\la_\al^k\:,\hs\hs n\geq2\:.
\nn\eeq

In the latter equation, the restriction on the range of $k$ 
can be easily derived by dimensional considerations. 
Using the above results, for the quantities we need in the paper,
after straightforward calculations we get
\beq 
\La_{20}&=&\frac16\lap\at\frac{R}{6}-V\ct\:,
\\
e^{-2\si}\La_{21}&=&-\frac13\lap\si-\frac23\si^k\si_k\:,
\\
e^{-2\si}\La_{31}&=&-\frac13\si^k\nabla_k\at\frac{R}{6}-V\ct
-\frac1{90}\si^k\nabla_kR
-\frac1{45}R^{ij}(7\si_{ij}+2\si_i\si_j)
\nn\\&&\hs
-\frac1{30}\lap^2\si-\frac1{15}(\lap\si)^2
-\frac4{15}\si^k\si_k\lap\si
\nn\\&&\hs
-\frac1{15}\aq4(\si^k\si_k)^2+4\si^k\nabla_k\lap\si
+2\si^{ij}\si_{ij}+8\si^{ij}\si_i\si_j\cq\:,
\\
e^{-4\si}\La_{32}&=&\frac13\si^k\si_k
\\
e^{-4\si}\La_{42}&=&\frac1{15}R^{ij}\si_i\si_j
+\frac1{18}(\lap\si)^2+\frac{14}{15}(\si^k\si_k)^2
+\frac{22}{45}\si^k\si_k\lap\si
\nn\\&&\hs
+\frac2{15}\si^k\nabla_k\lap\si+\frac2{45}\si^{ij}\si_{ij}
+\frac{32}{45}\si^{ij}\si_i\si_j\:,
\\
e^{-6\si}\La_{53}&=&-\frac19\si^k\si_k\lap\si
-\frac{22}{45}(\si^k\si_k)^2-\frac{2}{15}\si^{ij}\si_i\si_j\:,
\\
e^{-8\si}\La_{64}&=&\frac1{18}(\si^k\si_k)^2\:.
\eeq


\begin{thebibliography}{10}


\bibitem{wipf}
V. Mukhanov, A. Wipf and A. Zelnikov.
Phys. Lett. {\bf B 332}, 283(1994).

\bibitem{eli}
E.~Elizalde, S. Naftulin and S. D. Odintsov.
Phys.~Rev. {\bf D49}, 2852 (1994);
S. Nojiri and S. D. Odintsov.
Phys. Lett {\bf B463}, 57 (1999).


\bibitem{frolov99}
V. Frolov, P. Sutton and  A. Zelnikov. 
Phys. Rev. { D 61}, 02421, (2000).

\bibitem{sutton}
P. Sutton. 
Phys. Rev. { D 62}, 044033, (2000).

\bibitem{balbinot00}
R.~Balbinot, A.~Fabbri, V.~Frolov, P.~Nicolini, P.~Sutton, A.~Zelnikov,
{it Vacuum polarization in the Schwarzschild spacetime 
and dimensional reduction}. hep-th/0012048.

\bibitem{eliz94b}
E.~Elizalde, S.~D.~Odintsov, A.~Romeo, A.A.~Bytsenko and S.~Zerbini.
{\em Zeta Regularization Techniques with Applications}.
World Scientific, Singapore (1994).

\bibitem{byts96-266-1}
A.A.~Bytsenko, G.~Cognola, L.~Vanzo and S.~Zerbini.
 Phys.~Rep. {\bf 266}, 1 (1996).


\bibitem{eliz98-194-613}
E.~Elizalde, L. Vanzo and S. Zerbini.
 Commun.~Math.~Phys. {\bf 194}, 613 (1998).

\bibitem{eliz98-57-7430}
E.~Elizalde, A. Filippi, L. Vanzo and S. Zerbini.
  Phys.~Rev. {\bf D57}, 7430 (1998).

\bibitem{eliz98-532-407}
E.~Elizalde, G.~Cognola and S. Zerbini.
 Nucl.~Phys.  {\bf B532}, 407 (1998).


\bibitem{ray73-98-154}
D.B.~Ray and I.M.~Singer.
Ann. Math. {\bf 98}, 154 (1973).

\bibitem{hawk77-55-133}
S. W. Hawking.
Commun. Math. Phys. {\bf 55}, 133 (1977).

\bibitem{dowk76-13-3224}
J.S.~Dowker and R.~Critchley.
 Phys.~Rev. {\bf D 13}, 3224 (1976).


\bibitem{kirsten}
 G.~Cognola, K. Kirsten and S. Zerbini.
 Phys. Rev.  {\bf D 48}, 790 (1993).


\bibitem{dewi65b}
B.S.~DeWitt.
{\em The Dynamical Theory of Groups and Fields}.
Gordon and Breach, New York (1965).

\bibitem{seel67-10-172}
R.T.~Seeley.
 Am.~Math.~Soc.~Prog.~Pure Math. {\bf 10}, 172 (1967).


\bibitem{parker}
L. Parker and D.J. Toms.
Phys. Rev {\bf D31 }, 953 (1985); Phys. Rev {\bf D31 }, 953 (1985).


\bibitem{moretti}
D. Iellici and V. Moretti. 
Phys.Lett. {\bf B435},33 (1998).

\bibitem{binosi}
D. Binosi and S. Zerbini. 
 J.Math.Phys. {\bf 40}, 5106 (1999).

\bibitem{camporesi}
R. Camporesi.
Phys. Rep. {\bf 196}, 1 (1990).

\bibitem{bcz96}
A. A. Bytsenko,G. Cognola and S. Zerbini.
Nucl.  Phys. {\bf B458}, 267 (1996).
 

\bibitem{all}
R. Bousso and S. Hawking.
Phys. Rev.  {\bf D56}, 7788 (1997);
T. Chiba and M. Siino.
Mod. Phys. Lett.  {\bf A 12}, 709 (1997);
W. Kummer, H. Lieb  and D.V. Vassilevich.
Mod. Phys. Lett.  {\bf A 12}, 2683 (1997).

\bibitem{all1}
S. Nojiri and S. D. Odintsov.
Mod. Phys. Lett. {\bf A 12}, 2083 (1997);
S. Nojiri and S. D. Odintsov.
Phys. Rev. {\bf D57}, 2363 (1998);
S. Nojiri and S. D. Odintsov.
Phys. Rev. {\bf D57}, 4847 (1998);
S.J. Gates,S. Nojiri,T. Kadoyosi and S. D. Odintsov.
Phys. Rev. {\bf D58}, 084026 (1998);
W. Kummer, H. Lieb  and D.V. Vassilevich.
Phys. Rev.  {\bf D58}, 108501 (1998);
S. Nojiri, O. ObregonS. D. Odintsov and K. E. Osetrin.
Phys. Rev. {\bf D60}, 0204008 (1999);
R. Balbinot and A. Fabbri
Phys. Rev.  {\bf D59}, 044031 (1999);
R. Balbinot and A. Fabbri
Phys. Lett.  {\bf B459}, 112 (1999);
R. Balbinot, A. Fabbri and I. Shapiro.
Phys. Rev. Lett.   {\bf 83}, 1494 (1999);
R. Balbinot, A. Fabbri and I. Shapiro.
Nucl. Phys.   {\bf B559}, 301 (1999);
W. Kummer, H. Lieb  and D.V. Vassilevich.
Phys. Rev.  {\bf D60}, 084021 (1999);
F. C. Lombardo, F. D. Mazzitelli  and J. G. Russo.
Phys. Rev. {\bf D59}, 064007 (1999).
R. Balbinot, A. Fabbri, V. Frolov, P. Nicolini, P. Sutton, A. Zelnikov.
hep-th/0012048.

\end{thebibliography}
\end{document}